\documentclass[12pt]{article}

\begin{document}
\baselineskip 0.6cm

\newcommand{\defi}{\stackrel{\triangle}{=}}

\baselineskip 0.6cm
\newcommand{\gsim}{ \mathop{}_{\textstyle \sim}^{\textstyle >} }
\newcommand{\lsim}{ \mathop{}_{\textstyle \sim}^{\textstyle 3<} }
\newcommand{\vev}[1]{ \left\langle {#1} \right\rangle }
\newcommand{\bra}[1]{ \langle {#1} | }
\newcommand{\ket}[1]{ | {#1} \rangle }
\newcommand{\Dsl}{\mbox{\ooalign{\hfil/\hfil\crcr$D$}}}
\newcommand{\nequiv}{\mbox{\ooalign{\hfil/\hfil\crcr$\equiv$}}}
\newcommand{\nsupset}{\mbox{\ooalign{\hfil/\hfil\crcr$\supset$}}}
\newcommand{\nni}{\mbox{\ooalign{\hfil/\hfil\crcr$\ni$}}}
\newcommand{\EV}{ {\rm eV} }
\newcommand{\KEV}{ {\rm keV} }
\newcommand{\MEV}{ {\rm MeV} }
\newcommand{\GEV}{ {\rm GeV} }
\newcommand{\TEV}{ {\rm TeV} }

\newcommand{\be}[1]{\begin{equation}\label{#1}}
\newcommand{\ee}{\end{equation}}
\newcommand{\ds}{\displaystyle}
\newcommand{\bea}[1]{\begin{eqnarray}\label{#1}}
\newcommand{\eea}{\end{eqnarray}}
\newcommand{\ra}{\rightarrow}

\def\diag{\mathop{\rm diag}\nolimits}
\def\tr{\mathop{\rm tr}}

\def\Spin{\mathop{\rm Spin}}
\def\SO{\mathop{\rm SO}}
\def\O{\mathop{\rm O}}
\def\SU{\mathop{\rm SU}}
\def\U{\mathop{\rm U}}
\def\Sp{\mathop{\rm Sp}}
\def\SL{\mathop{\rm SL}}
\def\Qt{{\tilde Q}}
\def\mt{\tilde{m}}
\def\change#1#2{{\color{blue}#1}{\color{red} #2}\color{black}\hbox{}}


\begin{titlepage}

\vskip 2cm
\begin{center}
{\large \bf  Two Dimensional (0,2) Theories and Resolved $A_n$ singularities}
\vskip 1.2cm
Yizhuo Gao and Radu Tatar 

\vskip 0.4cm

{\it Division of Theoretical Physics, Department of Mathematical Sciences

The University of Liverpool,
Liverpool,~L69 3BX, England, U.K.

yizhuo.gao@liverpool.ac.uk,~~rtatar@liverpool.ac.uk}

\vskip 1.5cm

\abstract{We study (0,2) two-dimensional theories in type IIB configurations with D5 branes wrapping  
blow-up ${\bf{P}}^1$ cycles of deformed resolutions for $A_n$ singularities or in T-dual IIA configurations with suspended D4 branes. We consider deformations of four dimensional ${\cal{N}}=2, \prod_{i=1}^{n} U(N_i)$ theories with general superpotentials for the adjoint and bifundamental fields together with fundamental flavours and reduce to two dimensions on a two torus in the presence of magnetic fluxes and  FI terms.} 

\end{center}
\end{titlepage}


\section{Introduction}

In recent years, a sustained effort has been dedicated to describing theories with the lowest amount of supersymmetry, as are the two-dimensional (0,2) theories. Constructions were developed using  brane brick models \cite{franco1}-\cite{franco3}, 
Spin(7) orientifolds \cite{franco4}, F theory compactifications on Calabi-Yau 5-folds \cite{sakura1}-\cite{sakura3} and D1 branes at Calabi-Yau 4-folds singularities \cite{closset1}.

One very interesting approach was proposed in \cite{kl}, \cite{kl1}, using D4 branes 
suspended between NS branes when magnetic fluxes $B$ and FI-terms are turned on the D4 branes. When starting with  ${\cal{N}}=1$ theory in four dimensions on suspended D4 branes, by further wrapping D4 branes on a two torus with magnetic flux and turning on a D-term, supersymmetry is partially preserved when the magnetic flux equals the D term and the resulting theory becomes (0,2) in the two uncompactified directions. When applying a  T-duality, the suspended D4 branes are mapped to D5 branes wrapped on ${\bf P}^1$ cycles. A D term corresponds to a 
nonzero volume for the  ${\bf P}^1$ cycles which would break supersymmetry when two stacks of D5 branes are wrapped on  ${\bf P}^1$ cycles with different volumes. A further $T^2$ compactification with different magnetic fluxes on the $T^2$ bundle over the two  ${\bf P}^1$ cycles can lead to a (0,2) theory in 2 dimensions.

Some further steps in this direction were considered in \cite{tatar2015} and \cite{tatar2017} for gauge group $U(N_c)$ with 
$N_f$ fundamental flavours. The field theory lives on $N_c$ D5 branes wrapping a ${\bf P}^1$ cycle of finite size and $N_f$ D5 branes wrapping a  ${\bf P}^1$ cycle of very large/infinite size. In case of even $N_f$, after choosing a  $U(1) $ inside the global symmetry $U(N_f)$ and giving charges $e$ to half of the flavours and $-e$ to the other half, the flavour group is broken to $SU(N_f/2) \times SU(N_f/2)$ and corresponds to a realignment of  ${\bf P}^1$ cycles in space time directions corresponding to nonzero FI term $D$. To preserve supersymmetry, a further compactification on $T^2$ is required with magnetic flux $B$. The condition $B=D$ becomes a partial supersymmetry preservation and is equivalent to a requirement for  D5 branes to wrap holomorphic 4 cycles.
\cite{tatar2017} considered (0,2) two-dimensional theories arising from branes wrapped on deformations and resolutions of $A_1$ singularity leading to ${\cal{N}}=1$ in four dimensions. The path from  branes wrapped on resolved $A_1$ singularity in four dimensions to
 ${\cal{N}}=1$ theories in four dimensions was detailed in 
\cite{ot} and the further step from ${\cal{N}}=1$  to (0,2) in two dimensions was considered in \cite{tatar2017}.

There are two possible paths to descend from $4d~{\cal{N}}=2$ theories to (0,2) theories in two dimensions depending on choosing the order of turning on the magnetic flux and D term or adding the superpotential for adjoint fields. 

\vskip .6cm
 
We can turn on first the superpotential for the adjoint fields and then the magnetic flux and the path is

$$1.~4d~{\cal{N}}=2~\rightarrow~4d~{\cal{N}}=1~\rightarrow~2d~(0,2)$$

We can turn on the magnetic flux first and then the superpotential and the path is 

 $$2. ~4d~{\cal{N}}=2 \rightarrow 2d~(2,2) \rightarrow 2d (0,2)$$
 
As discussed in \cite{tatar2017}, as $4d~{\cal{N}}=2$ becomes  (4,4) in two dimensions, by turning on equal magnetic flux and FI can break the (4,4) supersymmetry to (2,2) but not to (0,4) or (4,0).  The field theory lives on wrapped D5 on holomorphic 4-cycles. The step $2d~(2,2) \rightarrow 2d (0,2)$ follows from adding a superpotential and considering deformations of the geometry with 4-cycles.

The $4d~{\cal{N}}=2$ and $2d~(2,2) $ theories have a chiral field $\Phi$  in the adjoint representation which can have a superpotential with arbitrary powers of $\Phi$. Two important cases discussed in \cite{ot} considered quadratic superpotential $\mbox{Tr} \Phi^2$ or degenerate superpotential $\mbox{Tr} \Phi^k$. In the current work, we consider mostly the quadratic case and leave the general case for future study. We discuss deformations of ${\cal{N}} = 2$ theories in 4 dimensions on $n$ stacks of $N_i, i=1,\cdots,n$  D5 branes wrapped 
on the resolution $i-th$ ${\bf P}^1$ cycle. The field theory when all the ${\bf P}^1$ cycles are of finite size is   $\prod_{i=1}^{n} U(N_i)$ with bifundamentals.  This changes if we choose one of the ${\bf P}^1$ cycles to be very large. If ${\bf P}^1_1$ is very large, $U(N_1)$ becomes a flavour group, and the $U(N_2)$ gauge group has $N_1$ fundamental flavours. If ${\bf P}^1_n$ is very large, $U(N_n)$ becomes a flavour group, and $U(N_{n-1})$ gauge group has $N_n$ fundamental flavours. If ${\bf P}^1_k, 1<k<n$ is very large, $U(N_k)$ becomes a flavour group and both $U(N_{k-1})$ and $U(N_{k+1})$ gauge groups have $N_k$ fundamental flavours.For all the choices of flavour
fundamental $U(N)$ group, we consider the Abelian part $U(1)$ of the global group and turn on the magnetic flux and the FI terms, leading to a large set of (0,2) theories in two dimensions.

In section 2 we review some of the results of \cite{ot} regarding the deformation of $4d~{\cal{N}}=2$ theories on D5 branes wrapped on blow-up cycles of $A_n$ singularities or their T-dual picture with D4 branes on intervals between NS branes. We distinguish between finite intervals (corresponding to gauge groups) and long intervals (corresponding to flavour groups) and discuss the four dimensional theories of interest, with new geometric deformations induced by nonzero D terms. In section 3 we compactify the four dimensional theories to two dimensional theories in the presence of magnetic flux.In section 4 we add the superpotential for the adjoint field and discuss the corresponding (0,2) theories.

\section{Deformations of $A_n$ singularities}

\subsection{$\it{N} = 2$ theories in 4 dimensions}

We start by revising the set-up of \cite{ot}. Consider $A_n$ singularities embedded in $C^3$ as hypersurfaces

\begin{equation}
x y + u^{n+1} = 0.
\end{equation}

Their resolutions involve a collection of $n~{\bf P}^1$ cycles together with their normal bundles $O(-2) \oplus O(0)$. For each 
$O(-2)$ over ${\bf P}^1$ , the map between North pole coordinates $Z'_{i}, Y'_{i}$ and  and South pole coordinates  $Z_{i}, Y_{i}$ is 
\begin{equation}
Z'_{i} = 1/Z_i,~~Y'_{i} = Y_i Z_{i}^2.
\end{equation}
The total space is obtained by gluing a neighbourhood of the North pole ($Z'_i=0, Y'_i=0, X'_i= 0$) of ${\bf P}^1_i$ and one of the South pole ($Z_{i+1}=0, Y_{i+1}=0, X_{i+1}= 0$) of ${\bf P}^1_{i+1}$.

 This is an isomorphism induced by the map $Y'_i \rightarrow Z_{i+1},~Z'_i \rightarrow Y_{i+1}$. The circle action
\begin{eqnarray}
(e^{i \theta},~Z_i) & = & e^{i \theta} Z_i,~~ (e^{i \theta},~Y_i) = e^{- i \theta} Y_i,  \\ \nonumber
(e^{i \theta},~Z'_i) & = & e^{-i \theta} Z'_i,~~ (e^{i \theta},~Y'_i) = e^{i \theta} Y'_i, 
\end{eqnarray} 
is compatible with the plumbing as it has the same action on the pairs $Z'_i, Y_{i+1}$ and $Y'_i, Z_{i+1}$ respectively.The orbits of the action degenerate along $Z_i = Y_i =0$ and $Z'_i = Y'_i =0$ which signal the presence of parallel NS branes in the T dual picture.  

For an $A_n$ singularity, the T dual picture contains $n+1$ NS branes that we denote $NS_0$ to $NS_n$ from left to right.  If we want the NS branes to be separated along the radial direction of the  ${\bf P}^1$ cycles,
the distance between $NS_i$ and $NS_{i+1}$ is exactly given by some integral of the NS field $B_{NS}$ on ${\bf P}^1_i$,
$b_i = \int_{{\bf P}^1_i} B_{NS}$. If we wrap $N_i$ D5 branes on each of the   ${\bf P}^1_i$ cycle $i=1,\cdots,n$, we get some ${\cal{N}} = 2$ theories in 4 dimensions with the gauge group  $\prod_{i=1}^{n} U(N_i)$. The D5 branes are mapped under T-duality to D4 branes on the segments between NS branes. Consider that the parallel NS branes are extended along the (012345) directions and the D4 branes are along the (01236) directions, with $x^6$ being the intervals between the NS branes.

The bare gauge coupling constants $g_i$ for the $U(N_i)$ groups are given by
\begin{equation}
\frac{4 \pi}{g_i^2} =  \frac{b_i}{g_s}
\end{equation}
The coupling constants $g_i$ can be varied by changing $b_i$ i.e. by changing $B_{NS}$ through ${\bf P}^1_i$. If $b_i$ is large enough, $g_i$ can be considered near zero and the corresponding gauge group $U(N_i)$ becomes a global group. In the T-dual picture, this corresponds to the length of one interval being very big. 

We consider two cases when  an interval becomes very large:

(a) the long interval is at the very right or at the very left.. That means that the $b_2,\cdots,b_{N-1}$ are all finite, whereas
either $b_1$ or  $b_n$ is very large. This implies that either  
$\prod_{i=1}^{n-1} U(N_i)$ is a gauge group whereas $U(N_n)$ is a global group or that $\prod_{i=2}^{n} U(N_i)$ is a gauge group and $U(N_1)$ is a global group  . We will discuss the case of $U(N_n)$  global group which implies that the long interval is at the right. This implies that the group $U(N_{n-1})$ has some extra fundamental flavors $F_{i}, i=1,\cdots,N_{n}$. The abelian part of $U(N_{n})$ is the $U(1)$ group whose D term we are interested and we will discuss the D term and magnetic flux later on.  

(b) one of the other intervals is very long. Consider that the interval $1 <  k < n$ is very long so $U(N_k)$ is a candidate for the global group whereas $\prod_{i=1, i\ne k}^{n} U(N_i)$ is a gauge group.In this case the gauge groups $U(N_{k-1})$ and 
$U(N_{k+1})$ both have extra $N_k$ fundamental flavours $G_{i}, i = 1,\cdots, N_{k}$ . The gauge group $U(N_i),~i \ne k-1,k,k+1$ are unchaged.

\subsection{${\cal{N}} = 1$ theories in 4 dimensions}

Deform now the ${\cal{N}} = 2$ theories by adding a tree-level superpotential which takes the form
\begin{equation}
W = \sum_{i=1}^{n} \mbox{Tr} (\sum_{j=1}^{d_i + 1}\frac{g_{i,j-1}}{j} \Phi_i^j -  Q_{i,i+1} \Phi_{i} Q_{i+1,i})
\end{equation}
where $Q_{i,i+1}$ are bifundamental  fields in the $(N_{i}, \bar{N}_{i+1})$ representation of the group $U(N_i) \times U(N_{i+1})$ and 
 $Q_{i+1,i}$ are fields in the $(\bar{N}_{i}, N_{i+1})$ representation of the group $U(N_i) \times U(N_{i+1})$. 

As in \cite{ot}, the two cases with the clearest geometric interpretation are the quadratic case

\begin{equation}
W = \mbox{Tr} (\sum_{i=1}^{n}( \frac{g_i}{2} \Phi_i^2 + h_i \Phi_i - \mbox{Tr} Q_{i,i+1} \Phi_{i} Q_{i+1,i})
\end{equation}

and the degenerate superpotential 

\begin{equation}
W = \mbox{Tr} (\sum_{i=1}^{n}( \frac{g_i}{m+1} \Phi_i^{m+1}  - \mbox{Tr} Q_{i,i+1} \Phi_{i} Q_{i+1,i})
\end{equation}

\subsection{Quadratic Superpotential}

For the quadratic case, the vevs of the bifundamentals fields break the gauge group from $\prod_{i=1}^{n} U(N_i)$  to 

\begin{equation}
\prod_{i=1}^{n} U(M_{i}) \times \prod_{i=1}^{n-1} \prod_{k=i}^{n-1} U(M_{i,k})
\end{equation}
where the group $U(M_{j,k})$ is embedded diagonally in $\prod_{i=j}^{k} U(N_i)$. 

For $n=2$, the  ${\cal{N}} = 2$ has gauge group $U(N_1) \times U(N_2)$ . For the quadratic deformation
\begin{equation}
W = \mbox{Tr} (\frac{g_1}{2} \Phi_1^2 + \frac{g_1}{2} \Phi_2^2  + h_1 \Phi_1 +  h_2 \Phi_2 ) - \mbox{Tr} (Q_{1,2} \Phi_{2} Q_{2,1})
-  \mbox{Tr} (Q_{2,1} \Phi_{1} Q_{1,2}) ,
\end{equation}
the vevs of the $Q_{1,2} Q_{2,1}$ break the gauge group to 
$U(M_1) \times U(M_2) \times U(M_{1,2})$, where $M_1, M_2$ correspond to the zero vevs and $M_{1,2}$ to the nonzero vevs. $U(M_{1,2})$ is diagonally embedded into $U(N_1) \times U(N_2)$ . The values of $M$ and the values of  $N$ are related by 
$M_1 + M_{1,2} = N_1, M_2 + M_{1,2} = N_2$. 

For $n=3$,  ${\cal{N}} = 2$, the gauge group $U(N_1) \times U(N_2) \times U(N_3)$  is broken by the nonzero vevs for 
$Q_{1,2} Q_{2,1}$,~$Q_{3,2} Q_{2,3}$, ~$Q_{1,2} Q_{2,3} Q_{3,2} Q_{2,1}$  to $\prod_{i=1}^3 U(M_i) \times U(M_{1,2}) \times U(M_{1,3}) \times U(M_{2,3})$, where $M_1, M_2, M_3$ correspond to zero vevs,
$M_{1,2}$ to nonzero  $Q_{1,2} Q_{2,1}$, $M_{2,3}$ to nonzero  $Q_{2,3} Q_{3,2}$ and $M_{1,3}$ to nonzero $Q_{1,2} Q_{2,3} Q_{3,2} Q_{2,1}$. 

We have  $M_1 + M_{1,2}  + M_{1,3}= N_1,  M_2 + M_{1,2} + M_{2,3}+ M_{1,3}= N_2, M_3 + M_{2,3} + M_{1,3} = N_3$.

How does the discussion change when one of the gauge groups becomes a global group? We consider the two cases mentioned before:

(a) $U(N_1)$ or $U(N_n)$ become  global groups. The bifundamental fields $Q_{n-1,n}$ and $Q_{n,n-1}$ are now fundamental flavours for the gauge group $U(N_{n-1})$ or  $Q_{1,2}$ and $Q_{2,1}$ are now fundamental flavours for the gauge group $U(N_{2})$ . We only 
give the details for the $U(N_n)$ global group. 

For $n=2$ the gauge group is $U(N_1)$ with $N_2$ fundamental fields.If $P$ of the fundamental fields have a  vev, the 
gauge group is broken to $U(N_1 - P)$ and the flavour group becomes $U(N_2 - P) \times U(P)$ .

For $n=3$ the gauge group is $U(N_1) \times U(N_2) $ with $N_3$ fundamental flavours for $U(N_2)$ group.  After Higgsing, the gauge group is 
$U(M_1) \times U(M_2) \times U(M_{1,2})$ where $U(M_{1,2})$ corresponds to vevs for $Q_{1,2} Q_{2,1}$. The groups $U(M_3), U(M_{1,3}), U(M_{2,3})$ becomes global groups,  
$U(M_3)$ correspond to zero vevs for the bifundamentals   $Q_{1,2} Q_{2,1}$ and fundamentals $Q_{2,3} Q_{3,2}$, 
$U(M_{2,3})$ to  zero vevs for the bifundamentals   $Q_{1,2} Q_{2,1}$ but nonzero for fundamentals $Q_{2,3} Q_{3,2}$,
 $U(M_{1,3})$ to  nonzero vevs for both bifundamentals   $Q_{1,2} Q_{2,1}$ and fundamentals $Q_{2,3} Q_{3,2}$. The remaining gauge group is 
$U(M_1) \times U(M_2) \times U(M_{1,2})$ with $M_3$ fundamental flavours for $U(M_2)$.

(b) the interval $1 <  k < n$ is very long and $U(N_k)$ is a global group. The bifundamental fields $Q_{k-1,k}$ and $Q_{k,k-1}$ are now fundamental flavours for the gauge group $U(N_{k-1})$ whereas  the bifundamental fields $Q_{k,k+1}$ and $Q_{k+1,k}$ are now $N_k$ fundamental flavours for the gauge group $U(N_{k+1})$.

For $n=3$, the gauge group is $U(N_1) \times U(N_3)$ with $N_2$ fundamental flavours for both groups and no bifundamental flavours. If we give a vev to $m$ fundamental flavours, the gauge group becomes $U(M_1) \times U(M_3)$ with $M_2$ fundamental flavours where $M_i = N_i - m, i=1,2,3$. 

For $n=4$, we take the gauge group  $U(N_1) \times U(N_3) \times U(N_4)$ with $N_2$ fundamental flavours for 
$U(N_1)$ and  $U(N_3)$, which corresponds to the second interval being very long. The only bifundamental fields are 
 $Q_{3,4}$ and $Q_{4,3}$. If giving expectation values to $m$ fundamental flavours, the gauge group
becomes $U(N_1 -m) \times U(N_3-m) \times U(N_4)$ as $U(N_4)$ is unaffected. If, in addition, we give vev to $m'$ bifundamental fields, the gauge group becomes $U(N-1 - m) \times U(N_3-m - m') \times U(N_4 - m') \times U(m')$ as $U(N_1 - m)$ is now unaffected
and $U(m')$ is a gauge group.  The 2 stacks of  D4 branes to the right of the long interval behave as if they come from a deformed $A_2$ singularity.

For $n=5$, if the second interval is long, the stacks to its right behave like a deformed $A_3$ singularity. If the third interval is long,
the stacks to its left behave like a deformed $A_2$ singularity and the same for the stacks to its right.

We can also proceed further and make some further gauge groups global. For $n=4$, could start from $U(N_1 -m) \times U(N_3-m) \times U(N_4)$  and make any of $U(N_1 -m),  U(N_3-m)$ or  $U(N_4)$  global. If $U(N_1 -m)$ becomes global,
 the gauge group remains $U(N_3-m) \times U(N_4)$ with extra $N_1-m$ fundamental flavours for $U(N_3-m)$.   If $U(N_4)$ becomes global,
 the gauge group remains $U(N_1-m) \times U(N_3-m)$ with extra $N_4$ fundamental flavours for $U(N_3-m)$.  If $U(N_3 -m)$ becomes global 
 the gauge group as two separated  $U(N_1-m)$ and $U(N_4)$ groups with extra $N_3-m$ fundamental flavours for both groups. As $n$ increases, we could have several other flavour groups.

\subsection{The "flavour" Interval}

Consider four NS branes parallel in the (012345) directions and denote them by A,B, C, and D. Add $N_1$ D4 branes between A and B and $N_3$ branes between C and D and leave the interval between B and C empty. The fundamental flavours for $U(N_1)$ are introduced as 
$N_2$ D6 branes on (01234789) directions between NS branes A and B, intersecting the $N_1$ D4 branes.  The fundamental flavours for $U(N_3)$ could be introduced as  $N_2$ D6 branes on (01234789) directions between NS branes C and D, intersecting the $N_3$ D4 branes.

Now we move the left stack of D6 branes to the right. When it crosses the NS brane B, the Hanany Witten effect  \cite{hw} tells us that D4 branes are created between the left  D6 branes and the NS brane A. As the left D6 branes move to the right, the created D4 branes fill the left part of the interval. If we move the right stack of D6 branes to the left, they will cross the NS brane C. A Hanany-Witten effect tells us that  D4 branes are created between the NS brane C and the right D6 branes as they move to the left of C, and they fill the right part of the interval. The question is what happens when the two stacks of D6 branes approach each other? If they pass each other, the D4 branes that were created could be extended, and there could be portions of the interval with $2 N_2$ D4 branes which would contradict the fact that we have $N_2$ D4 branes on the long interval. 

If we go to the geometrical picture by a T-duality, the D4 branes become D5 branes wrapped on ${\bf{P}}^1$ cycle, the NS branes are the lines of singularity discussed before, and the D6 branes become flavour D5 branes. For the ${\bf{P}}^1$ cycle with no wrapped D5 branes, moving the flavour D5 branes across the lines of singularity leads to creating D5 wrapped on the corresponding  ${\bf{P}}^1$ cycle. As in the brane configuration, we consider that the D5 branes crossing the lines of singularity from different directions can't cross each other. It would be interesting to understand this in more detail.

 \subsection{Degenerate Superpotential and its deformations}

For the degenerate superpotential 
\begin{equation}
W = \mbox{Tr} (\sum_{i=1}^{n}( \frac{g_i}{m+1} \Phi_i^{m+1}  - \mbox{Tr} Q_{i,i+1} \Phi_{i} Q_{i+1,i})
\end{equation}
we know from \cite{ot} that the deformation of $A_n$ singularity has $n(n~m + m - 1)$ deformations out of which $m (n+1) n/2$ are 
normalisable and correspond to wrapped D5 branes or D4 branes on intervals and $m (n+1) n/2 - n$ are non-normalizable and correspond to expectations values of various flavours. $m$ determines the shape of curved NS branes.

If the superpotentials for $\Phi_i$ are denoted by $W_i(\Phi)$, the first NS brane is given by $u = W'_1(\Phi)$, the second one by $u =  W'_1(\Phi) +  W'_2(\Phi)$, the third one by $u =  W'_1(\Phi) +  W'_2(\Phi) + W'_3(\Phi)$. The intersection between the $(l-1)$-th NS brane and the 
$m$-th NS brane is given by 
\begin{equation}
W'_l(\Phi) + W'_{l+1}(\Phi) + \cdots +  W'_m(\Phi) = 0
\end{equation}  

For the case when we have one long interval, the number of normalisable deformations decreases, and we expect the result for the deformation of $A_n$ to be equal to the one for the deformation of $A_{n-1}$. We leave the full discussion for future work. 

\section{Two dimensional theories}

We now compactify the four dimensional field theories discussed in the previous section on a two torus in the directions $x^1,x^2$. 
We choose an Abelian group $U(1)$ which is orthogonal to the gauge group. This is a  $U(1)$  inside $U(N_n)$ if the long interval is the one at the right or $U(N_k),~k\ne 1, k$  if the long interval is the $k$-th one.
We turn on a constant magnetic field $F_{12} = M$ through the torus and an equal D-term for the $U(1)$.

\subsection{(2,2) Theories in 2 dimensions}

We start with the ${\cal{N}}=2$ theory in four dimensions. The flavour fields are
${\cal{N}}=2$ hypermultiplets which consist of an ${\cal{N}}=1$ chiral multiplet, $(Q, \phi_Q)$ and an antichiral ${\cal{N}}=1$
multiplet $(\tilde{Q}^{\dagger}, \phi_{\tilde{Q}}^{\dagger})$. If the four dimensional theory exists in $(x^0, x^1, x^2, x^3)$, the compactfication on the two torus  $(x^1,x^2)$ provides a  (4,4) theory in 2 dimensions.

We turn on a constant magnetic field $F_{12} = M$ through the torus and the equal D term for the Abelian $U(1)$ symmetry. From \cite{kl} we know that the  ${\cal{N}}=1$ chiral multiplet, $(Q, \phi_Q)$  provides $e_i$ (0,2) chiral superfields for fields with U(1) charges $e_i > 0$ and $|e_i|$ Fermi superfields for 
$e_i < 0$, all in the fundamental representation.  As discussed in \cite{tatar2017}, the reductions of ${\cal{N}}=2, d=4$ with D term on a $T^2$ with magnetic flux equal to the D term takes $(Q,\phi_Q)$ and $(\tilde{Q}^{\dagger}, \phi_{\tilde{Q}}^{\dagger})$ into  the (2,2) matter multiplets $\Phi_{Q}^{(2,2)}$ and $\Phi_{\tilde{Q}}^{(2,2)}$. 

To fulfill the anomaly-free constraint for a global $U(1)$ orthogonal to the gauge group, the charges need to satisfy 
$\sum_i e_i = \sum_i \tilde{e}_i = 0$ where $e_i$ are charges of $Q_i$ and $\tilde{e}_i$ of $\tilde{Q}_i$. Due to the superpotential 
$Q^{e=1} \Phi \tilde{Q}^{e=-1} +  Q^{e=1} \Phi \tilde{Q}^{e=-1}$, the global symmetry $SU(N_n)$ is broken to
$SU(N_n/2)_1 \times SU(N_n/2)_2$ or  $SU(N_k)$ is broken to
$SU(N_k/2)_1 \times SU(N_k
/2)_2$ .  

We now consider the deformations of the above models (a) and (b)  for $A_n$ resolutions.

(a) The long interval is at the end so $\prod_{i=1}^{n-1} U(N_i)$ is a gauge group whereas $U(N_n)$ is a global group. We consider turning on equal  D term $j_n$  and magnetic field for the abelian $U(1)$ inside $U(N_n)$. The remaining $SU(N_n)$ is broken to 
$SU(N_n/2) \times SU(N_n/2)$ which is a flavour symmetry for the gauge group $U(N_{n-1})$. 

In the brane configurations, the $n+1$-th NS brane is replaced by two  NS branes separated from the original one by a distance $j_n$ in the $x^7$ direction. The stack of $N_n$ D4 between the original  $n$-th NS brane and the original $n+1$-th NS is replaced by two stacks of 
$N_n/2$ NS branes rotated by angles $\mbox{tan}(\theta) = \pm \frac{j_n}{b_n}$. The angle $\theta$ is very small for large $b_n$.

In the geometrical picture, the D term $j_n$ corresponds to the integral of Kahler form on the ${{\bf{P}}}^1$ that $N_n$ D5 branes wrap.  The magnetic field $B_i$ for the abelian part should be equal to the D-term that splits the stack of wrapped  
D5 branes into two stacks of $N_f/2$ D5 branes, for positive and negative values of the charge $e_i$. We then have the $n$-th  
${{\bf{P}}}^1$  with nonzero $j_n$ attached to the $n-1$-th ${{\bf{P}}}^1$  with zero $j_{n-1}$. The central charges for the D5 branes wrapped on n-th and (n-1)-th ${{\bf{P}}}^1$ do not match. This case can be linked to the discussion in \cite{ab} where the resolution of $A_2$ singularity was considered with D-tems. The vacuum energy for two ${{\bf{P}}}^1$ cycles wrapped by $N_{n-1}$ and $N_{n}$ D5 branes (consider a small separation $a$) is 
\begin{equation}
\label{supo1}
V_{*} = N_{n-1} \frac{b_{n-1}}{2~\pi~g_s} + N_n \frac{\sqrt{b_n^2 + j_n^2}}{2~\pi~g_s}
+\frac{1}{4 \pi^2} N_{n-1} N_n \mbox{log}(\frac{a}{\Lambda_0} (1 - \mbox{cos} \theta_{n,n-1}))+ \cdots
\end{equation} 
where $\theta_{n,n-1}$ is the relative phase between the central charges $Z_{n-1}~ = i b_{n-1}$ and $Z_{n} = j_n + i b_n$. A non-zero value for $\theta_{n,n-1}$ signals supersymmetry breaking. 

To restore partially the supersymmetry, we also turn the magnetic flux over the $n$-th ${{\bf{P}}}^1$ on the directions $x^1, x^2$.   
As discussed in \cite{tatar2017}, the condition for the restoration of supersymmetry over the  $n$-th ${{\bf{P}}}^1$ is
\begin{equation}
j_n  A_{T^2} = b_n M_n
\end{equation}
where $A_{T^2}$ is the area of the $T^2$ fiber and $M_n$ is the flux on the $T^2$ fiber over the $n$-th ${{\bf{P}}}^1$ .
This ensures that the $T^2$ fiber over the $n$-th ${{\bf{P}}}^1$ gives rise to a holomorphic 4-cycle. 

A T-duality on one of the directions of the two torus (say the $x^2$ direction) takes the $M_n$ flux into a vector potential along the direction $x_2$ equal to $A_2 = M_n~x_1$. The vector potential is associated with a global $U(1)$  realised on $D4$ branes wrapped on the 
$n$-th ${{\bf{P}}}^1$  and rotated in the $(x^1, x^2)$ plane.This is where we can see the difference between the flavour $Q_i$ with charges $e_i$  and flavour $\tilde{Q}_i$ with charges  $\tilde{e}_i$. The corresponding $D4$ branes are rotated by angles with the same absolute value but with different signs. Before  T duality on $x^2$, the difference between the signs of the charges was not visible.

A T-duality on the angular direction of the ${{\bf{P}}}^1$  cycles take all the wrapped D5 branes to D4 branes on intervals between 
two parallel NS branes, except the $N_n$ ones wrapped on the  $n$-th ${{\bf{P}}}^1$ which has nonzero $j_n$.  If the global symmetry $SU(N_n)$ is broken to $SU(N_n/2)_1 \times SU(N_n/2)_2$ by the choice of the magnetic charges,  the $N_n$ D4 branes 
split into two stacks of $N_n/2$ rotated by an angle $\pm \mbox{arctan} (\frac{j_{n}}{b_{n}})$ which end on two NS branes displaces in the $x^7$ direction by $j_{n}$.Again the corresponding $D4$ branes are rotated by angles with the same absolute value but with different signs in the $(x^6, x^7)$ plane. An extra T-duality in $x^2$ direction gives D3 branes rotated in both $(x^1,x^2)$ plane and  
$(x^6, x^7)$ plane and branes with different charges rotated differently. Supersymmetry is preserved if the angles of rotation are the same in the $(x^1,x^2)$ plane and  
$(x^6, x^7)$ plane. 

(b)  the interval $1 <  k < n$ is very long and $U(N_k)$ is a global group. Consider turning on a  D term and a magnetic field along the $T^2$ fiber for the abelian $U(1)$ inside $U(N_k)$. The remaining $SU(N_k)$ is broken for two 2 sets of magnetic charges with opposite signs into $SU(N_k/2) \times SU(N_k/2)$ which is a flavour symmetry for the gauge groups $U(N_{k-1})$ and $U(N_{k+1})$.
If there is a nonzero D-term $j_k$, the supersymmetry is completely broken. The superpotential (\ref{supo1}) for this case takes the form
\begin{equation}
\label{supo2}
V_{*} =\sum_{i=k-1,k+1} N_{i} \frac{b_{i}}{2~\pi~g_s} + N_k \frac{\sqrt{b_k^2 + j_k^2}}{2~\pi~g_s}
+\frac{1}{4 \pi^2} \sum_{i=k-1,k+1}(N_{i} N_k \mbox{log}(\frac{a}{\Lambda_0} (1 - \mbox{cos} \theta_{i,k}))+ \cdots
\end{equation} 
where $\theta_{k-1,k}$ is the relative phase between the central charges $Z_{k-1}~ = i b_{k-1}$ and $Z_{k} = j_k + i b_k$ whereas .  $\theta_{k+1,k}$ is the relative phase between the central charges $Z_{k+1}~ = i b_{k+1}$ and $Z_{k} = j_k + i b_k$ .Non zero values for $\theta$ signal supersymmetry breaking. 

To have reduced supersymmetry, we also turn the magnetic flux over the $k$-th ${{\bf{P}}}^1$ on the directions $x^1, x^2$.   
As discussed in \cite{tatar2017}, the condition for the restoration of supersymmetry over the  $k$-th ${{\bf{P}}}^1$ is
\begin{equation}
j_k  A_{T^2} = b_k M_k
\end{equation}
where $A_{T^2}$ is the area of the $T^2$ fiber and $M_k$ is the flux on the $T^2$ fiber over the $k$-th ${{\bf{P}}}^1$ .
This ensures that the $T^2$ fiber over the $k$-th ${{\bf{P}}}^1$ gives rise to a holomorphic 4-cycle wrapped by D5 branes.

What happens if we take a T-duality on the angular directions of the   ${{\bf{P}}}^1$ cycles? The D5 branes wrapped on all 
${{\bf{P}}}^1_i, i \ne k$ cycles become D4 branes along $x^6$ direction. The D5 branes on ${{\bf{P}}}^1_k$ become multiple  stacks of D4 branes which split as follows:

$\bullet$ the $k$-th NS brane splits into three NS branes, one denoted by $NS^{1}_k$ remaining in the original position and the other two denoted by  $NS^{2}_k$ and $NS^{3}_k$ having a displacement by $j_k$ in $x^7$  direction. There are two stacks of $N_k/2$ D4 branes between the $k+1$-th NS brane and $NS^{2}_k$ and $NS^{3}_k$ respectively which corresponds to a flavour group $SU(N_k/2) \times SU(N_k/2)$ for 
$U(N_{k+1})$. They are both rotated in the $(x^4,x^7)$ plane.

$\bullet$  the $(k+1)$ NS brane splits into three NS brane, one $NS^{1}_{k+1}$ remaining in the original position, $NS^{2}_{k+1}$ and $NS^{3}_{k+1}$ being displaced by $j_k$ in $x^7$  direction. There are two stacks of $N_k/2$ D4 branes between the $k$ NS brane and $NS^{2}_{k+1}$ and $NS^{3}_{k+1}$ respectively which corresponds to a flavour group $SU(N_k/2) \times SU(N_k/2)$ for $U(N_{k})$. 

\section{(0,2) Theories in 2 dimensions}

In this section, we discuss the supersymmetry breaking to (0,2) theories in 2 dimensions which is realised if we add a superpotential for the adjoint fields 
$\Phi^{2,2}_i$ in the adjoint representations of $\prod_{i=1}^{n-1} U(N_i)$ for case (a) or $\prod_{i=1, i \ne k}^{n} U(N_i)$ for case (b). We consider the case when the superpotential is either quadratic 
\begin{equation}
\label{02supotquad}
W = \mbox{Tr} \sum_{i=1}^{n}( \frac{g_i}{2} (\Phi^{2,2}_i)^2 + h_i \Phi^{2,2}_i) 
\end{equation}
or general
\begin{equation}
\label{02supotgen}
W = \mbox{Tr} \sum_{i=1}^{n}( \sum_{j=0} \frac{g_{i,j}}{j+1} (\Phi^{(2,2}_i)^{j+1})
\end{equation}
From \cite{ot}, we know that  $A_n$ singularities deformed with a power $m+1$ for the adjoint field and other terms taken to zero has 
$d_n=\frac{m(n + 1)n}{2}$ normalisable deformations corresponding to gauge theories and $d_n -n$ non-normalisable deformations that are specified by vacuum expectation values. 

\subsection{Quadratic Deformations}
For the quadratic deformations (\ref{02supotquad}), $m=1$  so there are $(n+1)n/2$ normalisable deformations and 
$n(n-1)/2$ non-normalisable deformations.The non-normalisable deformations correspond to the vacuum expectation values of the bifundamental fields and their products  
\begin{equation}
Q_{12}, Q_{23}, \cdots Q_{n-1 n}, Q_{12} Q_{23}, Q_{12} Q_{23} Q_{34} \cdots, 
Q_{12} Q_{23} \cdots Q_{n-1 n}
\end{equation}
  What happens when one of the gauge groups becomes a flavour group? We need to consider the two cases mentioned before.

\subsubsection{The last interval is very long}    

In this case the group $U(N_n)$ is flavour group.The bifundamental fields $Q_{n-1 n}$ become fundamental flavour for 
$U(N_{n-1})$ group. By giving expectation values to $Q_{n-2 n-1}$ and also considering  non zero values for 
$Q_{n-2 n- 1}Q_{n-1 n} $, the resulting group is a flavour group for $U(N_{n-2})$ . The sequence of non-zero expectation values for 
$Q_{23} \cdots Q_{n-1 n}$ gives a flavour group for $U(N_1)$ whereas the non-zero  expectation values for 
$Q_{12} \cdots Q_{n-1 n}$ gives a flavour group separated from the gauge groups.

All in all, $n$ of the gauge groups become flavour groups so the number of gauge groups decreases to $n(n-1)/2$, the number of vacuum expectation values remains at $(n-1)n/2$ but there are $n$ extra flavour groups that we can choose to turn on D terms and magnetic fluxes.

As discussed in the previous section, for $n=3$ we have 3 gauge groups $U(M_1) \times U(M_2) \times U(M_{1,2})$ and 3 flavour groups $U(M_{1,3}), U(M_{2,3})$ and $U(M_3)$ for D5 branes on long 
${{\bf{P}}}^1$ cycles. We can now choose nonzero $j_{1,3}, j_{2,3}, j_3$ for  any of these long ${{\bf{P}}}^1$ cycles and non zero magnetic flux $\tilde{M}_i$ on the $T^2$ fibre them. We use $\tilde{M}_{1,3}, \tilde{M_{2,3}}, \tilde{M_3}$ for the magnetic fluxes to differentiate from the values of 
$M$ in the gauge groups. The condition for preserving supersymmetry is 
\begin{equation}
j_{1,3}  (A_{T^2})_{2,3} = b_{2,3} \tilde{M}_{2,3} 
\end{equation}
In brane configurations, we would have 4 NS branes $NS_1, NS_2, NS_3, NS_4$ which are displaced in the 89 plane by the vevs of  $Q_{12}, Q_{23}, Q_{12} Q_{23}$. $NS_4$ is far from the others in the $x^6$ direction so 
the D4 branes between $NS_1 - NS_4, NS_2-NS_4, NS_3 - NS_4$ are flavour branes. If there are $M_1$ D4 branes between 
 $NS_1 - NS_4$, $M_2$ D4 branes between $NS_2-NS_4$ and $M3$ D4 branes between  $NS_3 - NS_4$, we can have various cases.

$\bullet$  We turn D term and flux for $U(1)$ inside $U(M_3)$ with half of the flavours with positive charge and half with negative charge.This breaks $SU(M_3)$ to $SU(M_3/2)\times SU(M_3/2)$ which live on D4 branes extended between $NS_3$ and two copies of 
$NS_4$ displaced in positive and negative $x^7$ direction. This case is obtained directly from the (2,2) discussion of the previous section. This is because nothing is displaced in the (89) direction.

$\bullet$ Every other combination is new for the (0,2) theories. We can turn D term and flux for $U(1)$ inside $U(M_{1,3})$ and/or $U(M_{2,3})$ with half of the flavours having a positive charge and half having a negative charge. This gives rise to either 
$SU(M_{1,3}/2)\times SU(M_{1,3}/2), SU(M_{2,3}/2)\times SU(M_{2,3}/2)$ or $SU(M_{1,3}/2)\times SU(M_{1,3}/2) \times 
SU(M_{2,3}/2)\times SU(M_{2,3}/2)$.
The first  case corresponds to  D4 branes extended between $NS_1$ and two copies of 
$NS_4$ displaced in positive and negative $x^7$ direction, second to  D4 branes extended between $NS_2$ and two copies of 
$NS_4$ displaced in positive and negative $x^7$ direction and third to when both types of D4 branes are present.
We can also have all three types of D terms, the ones inherited from (2,2) theory and the ones genuine for (0,2) theories.  

\subsubsection{ The $1<k<n$-th interval is very long}

We need at least 3 intervals so the condition is $n \ge 3$ for this case to exist.

 For $n=3$, the deformed ${\cal{N}}=1$ four dimension theory is $U(N_1) \times U(N_3)$ with $N_2$ fundamental flavours for both gauge grups. In the brane configurations, there are four NS branes rotated in the 89 directions. The vevs for $N$ of the $N_2$ fundamental flavours break the gauge group to a $U(N_1-N) \times U(N_3-N)$ gauge group with $N_2-N$ fundamental flavours for both and a separate $U(N)$ flavour group. 

The global group is then $U(N) \times U(N_2-N)$. There are various candidates for the global $U(1)$ with D-term and magnetic flux, one $U(1)$ inside $U(N_2-N)$ and one inside $U(N)$. The case with D-term for the $U(1)$ inside $U(N_2-N)$  corresponds to splitting $NS_2$ and $NS_3$ into 3 NS branes each, one 
in the original position and two with a displacement in the $x^7$ direction equal to the $D$ term. The (0,2) supersymmetry is preserved by choosing the appropriate magnetic field on the $(x^1,x^2)$ torus.   The case with D-term for the $U(1)$ inside $U(N)$  corresponds to splitting $NS_1$ and $NS_4$ into 3 NS branes each, one 
in the original position and two with a displacement in the $x^7$ direction equal to the $D$ term, together with appropriate magnetic fluxes to preserve supersymmetry.

What happens in the type IIB geometric picture? 

For $n=4$, if $k=2$, we start with  $U(N_1) \times U(N_3) \times U(N_4)$ gauge group with $N_2$ fundamentals $Q$ for 
 $U(N_1)$ and  $U(N_3)$ and bifundamentals $Q_{34}$ for  $U(N_3) \times U(N_4)$. We can give nonzero values to the flavour 
$Q, Q_{34}$ or to the combination $Q Q_{34}$. We consider that $N$ of the $N_2$ fundamental flavours $Q$, $N'$ of the bifundamental flavour $Q_{34}$ and $N''$ of the  $Q Q_{34}$ get a vev. The gauge groups becomes 
\begin{equation}
U(N_1 - N - N'') \times U(N_3 - N - N') \times U(N_4 - N')
\end{equation}

and the flavour groups is 

\begin{equation}
U(N) \times  U(N_2 - N - N'') \times U(N')
\end{equation}

 We see that now we have 4 choices for turning on $D$ terms and magnetic fluxes, first for $U(1)$ inside $U(N)$, second for 
$U(1)$ inside $U(N_2-N-N'')$, third for $U(1)$ inside $U(N')$ and fourth for $U(1)$ inside $U(N'')$  . The first choice would correspond to a split of the first and fourth NS brane into triplets of NS branes with a displacement equal to the D term for   $U(1)$ inside $U(N)$.  The second choice would correspond to a split of the second and third NS brane into triplets of NS branes with a displacement equal to the D term for   $U(1)$ inside 
$U(N_2-N-N'')$.  The third choice would correspond to a split of the third and fifth NS brane into triplets of NS branes with a displacement equal to the D term for $U(1)$ inside $U(N')$.   The fourth choice would correspond to a split of the first and fifth NS brane into triplets of NS branes with a displacement equal to the D term for $U(1)$ inside $U(N'')$.  In all these cases, an appropriate choice for the magnetic flux leads to (0,2) theories in two dimensions.

 For general $n > 4, k=2$, the gauge group is $U(N_1) \times U(N_3) \cdots \times U(N_n)$. There are $N_2$ fundamental flavours for 
$U(N_1)$ and $U(N_3)$ and bifundamental flavours $Q_{34}, Q_{45},\cdots,Q_{n-1 n}$ for the 
$U(N_3) \times U(N_4), U(N_4) \times U(N_5),\cdots, U(N_{n-1}) \times U_{n}$ respectively. We can then give expectation values to the combinations  
\begin{equation}
Q, Q_{34},Q_{45},\cdots, Q_{n-1~n}, Q Q_{34}, \cdots, Q Q_{ n-1~n}, Q Q_{34} Q_{45}, \cdots
\end{equation} 
and obtain a product of flavour groups. For each flavour group, we can turn a D term for its Abelian part which will correspond to a particular split of a set of NS branes into three NS branes each. A choice of equal magnetic fields leads to a large class of 
(0,2) theories. 

\subsection{General Superpotential Deformation}

We consider now the general deformation (\ref{02supotgen}). For $A_1$ singularity, the $\it{N}=1$ four-dimensional geometry is a union of two ${\bf{C}}^3$
($(x_1,y_1,z_1)$ and $(x_2,y_2,z_2)$ respectively) with the patching condition:
\begin{equation}
x_2 =1/ x_1,  y_2 = y_1,
 z_2 - z_1 x_1^2  = g_{m+1} \prod_{1}^{m} (x_1 x_2 - x_1 a_p)
\end{equation}

In the T-dual picture, this is a configuration with D4 branes suspended  between a straight NS brane and a curved one in the (4589) plane. There are $m$ stacks of D4 branes between the straight NS and the curved NS brane. 
For $A_n$ singularity, one gets curved NS branes in the directions $u = \sum_{i=0}^{p} W'_i(v)$, where $u=(x^8, x^9)$ and 
$v=(x^4, x^5)$. 

What happens when the interval between a pair of NS branes becomes large in ${\cal{N}}=2$ configurations? This would correspond to one of the gauge groups becoming global. By choosing one of the global groups to be a flavour group and singling out an $U(1)$ inside it, one can turn on a D term by displacing the NS branes in the $x^7$ direction. As the NS branes are now curved in the  (4589) plane, it is harder to envisage their displacement. This translates into a more involved geometries with holomorphic 4-cycles obtained after $T^2$ fibration with magnetic flux.  We will return to this more complicated model in a future publication.

\section{Conclusions}
In this work, we considered four-dimensional ${\cal{N}}=2, U(N_1) \times \cdots \times U(N_n)$ deformed to ${\cal{N}}=1$ theories originally discussed in \cite{ot}. The extra ingredient was a further compactification on a two torus with magnetic flux. Our consideration addresses the limit when one of the gauge groups becomes global with an FI term which is chosen to be equal to the magnetic flux in order to preserve half of the supersymmetry in two dimensions. We discussed some possible ways of choosing for the flavour group, each leading to a different class of (0,2) two-dimensional theories.

 Many other (0,2) two-dimensional theories could be constructed when some magnetic fluxes and FI terms are turned on.  \cite{rt} consider a large class of models involving orientifiolds, semi-infinite D4 branes, or deformations of complex structure which can lead to a large class of (0,2) theories with carefully chosen FI terms and magnetic fluxes. We will return to such models in a future publication.

\vskip 1cm

{\bf{Acknowledgments}}

The work of Radu Tatar was carried out with the support of the STFC Consolidated Grant ST/X000699/1. 
This research was supported in part by grant NSF PHY-2309135 to the Kavli Institute for Theoretical Physics (KITP).

\end{document}